# Laboratory investigation of nominally two-dimensional anabatic flow on symmetric double slopes

Roni H. Goldshmid[1,2,a)] and Dan Liberzon[1]

**Abstract**
We investigated the dynamics of highly turbulent thermally driven anabatic (upslope) flow on a physical model inside a large water tank using particle image velocimetry (PIV) and a thermocouple grid. The results showed that the flow exhibited pronounced variations in velocity and temperature and, importantly, could not be accurately modeled as a two-dimensional quasi-steady flow. Five significant findings are presented to underscore the three-dimensional nature of the flow. Namely, the B-shaped mean velocity profiles, B-shaped turbulent flux profiles, synthetic streaks that revealed particles flowing perpendicular to the laser sheet, average vorticity maps revealing helical structure splitting, and identified vortices shooting away from the boundary towards the apex plume. Collectively, these findings offer novel insights into the flow behavior patterns of thermally driven complex terrain flows, which influence local weather and microclimates and are responsible for scalar transport, e.g., pollution.

## I. INTRODUCTION

Complex mountainous topography characterizes more than 70% of urban areas and governs local weather and microclimates. Daily human activities are impacted by complex terrain weather, which imposes significant health, safety, and economic effects, e.g., pollution trapping or transfer in urban centers, formation of clouds, accurate weather forecasting, safety of civil and military aviation operations, sustainable and health aware urban planning, and operation of recreational centers.

Within mountainous regions, the flow characteristics of the local wind systems vary depending on the balance between thermal circulation flows (up/down-slope and up/down valley) and external flows (synoptic scale forcing). During the past decades substantial progress has been made in understanding synoptic scale flow approaching mountains and driven by large-scale pressure gradients[1–8], while significantly less progress has been made in understanding thermally driven flows. These flows result from thermal forcing due to diurnal solar heating/cooling cycle of slopes and are characterized as turbulent flows.

Turbulent slope flows that move upslope are known as anabatic flows, while the downslope ones are called katabatic flows. The relatively more stable conditions of katabatic flows, compared to anabatic flows, have led to a gap in the literature on anabatic flows[9], which we address in this work. Anabatic flows are driven by unstable stratification of air masses near the slope and tend to exhibit inhomogeneity, instability, and are non-stationary, especially in changing background field conditions at synoptic scales, e.g., mean wind speed and direction, temperature, and humidity[10]. Furthermore, the structure of the turbulent boundary layer (BL) and the flow separation behavior along the slope are of great importance. Together, they govern apex plume formation and entrainment of flow outside of the BL, which influence pollution transfer, cloud formation, and regional precipitation regimes[11–14].

Field and laboratory scale instruments, as well as data processing techniques, challenge the experimental investigation of highly turbulent and unstable anabatic flows. The current state of field work, as reflected in the available literature, primarily utilizes slow or insufficiently spatially resolved instruments, such as ultrasonic anemometers (sonic) or LiDAR (Light Detection and Ranging) sensors. These instruments are unable to capture fine-scale turbulent fluctuations, actual dissipation rates, mixing lengths, and heat transfer rates. Hot-wire/hot-film anemometers can capture fine scales but are impractical due to the constant recalibration requirement from changing field conditions and sensor degradation.

---

[1] Technion- Israel Institute of Technology, Faculty of Civil and Environmental Engineering
[2] Graduate Aerospace Laboratories, California Institute of Technology, Pasadena, CA 91125, USA
a) Author to whom correspondence should be addressed: ronig@caltech.edu

The combo probe is a recently developed instrument that consists of collocated sonic and hot-film sensors and addresses the calibration challenges associated with field measurements of turbulent flows[15–19]. It uses hot-film sensors to effectively sense and capture fine-scale turbulence, while the slower sonic records are used for continuous in-situ calibration of the hot-film (using neural networks) and for re-alignment along the mean-flow direction. A recent field experiment was conducted on a moderate slope of 5.7° reported on fine-scale turbulence statistics of thermally driven anabatic flow[17]. Point-measurement records were collected using the combo at 2 kHz, continuously for eight days. The analysis provided turbulence statistics of the flow, fully resolved spectra of velocity field components, and empirical correlations of various parameters to characterize the flow. To the best of our knowledge, this is the only field study to date that reports on the fine-scale turbulent characteristics of anabatic flows.

Field experiments are valuable, but modeling turbulent slope flows in the lab can provide additional insights. For example, laboratory experiments can be used to capture flow separation, quantify the transfer of pollutants, improve the space-time resolution of BL prediction models, and specify boundary conditions that require rapid spatial and temporal variability. The laboratory offers a highly controlled environment, which allows for higher accuracy measurements and repeatability and is particularly important for deriving empirical relations of various parameters governing the flow and for comparison with theoretical models[10] and numerical simulations[20,21].

Water tanks are commonly used in laboratories to model convective flows[22], slope flows[23–26], and other atmosphere-related flows[27–29]. A recent study explored the case of a uniformly smooth heated slope[25] while another study incorporated a plateau at the apex[26]. These studies empirically modeled separation location of the mean BL, mean upslope velocities at the point of separation, and their dependence on the slope angle, buoyancy flux, and plateau width. These empirical findings[25,26] provided a basis for the understanding of mean anabatic flow BL behavior but were not able to examine periodic flow patterns due to short runtimes of only few minutes determined by the tank volume. The experiment had to be stopped and restarted when the plume reached the slope toe. The short runtime prevented observation of periodic nature of the mean flow and a time to observe periodic mean behavior in the flow and limited the analysis to the first quasi-steady period. The BL remained stable until the plume started recirculating to the slope toe which resulted in the breaking of the BL. The flow was also found to be limited to two dimensions as the changes in the transverse axis appeared negligible, perhaps due to the finite size of the tank.

This work reports on a study conducted in a larger tank, which allowed a significant increase of the experiments' runtime, and allowed to examine increase the duration of the experiments and examine the steadiness of highly turbulent and potentially separating boundary layer flow. The results revealed that the flow in the larger tank was mainly governed by a three-dimensional behavior, which introduced a high level of intermittency in both mean and fluctuating parameters. The presented outcomes include the comparison of the separation location with previous models, BL characteristics, velocity profiles, temperature profiles, the baroclinic torque to vorticity advection ratio, and three-dimensionality of the flow. The combined results provide a comprehensive understanding of the flow patterns that are expected to be observed on a symmetric double slope mountain.

## II. EXPERIMENTAL SETUP

This laboratory study of thermally driven anabatic flow on a symmetric double slope was conducted at the Technion-Sea Atmosphere Interaction Research Laboratory (T-SAIL). The study used a large water tank (Figure 1) with dimensions of 2.0 m (length) by 0.6 m (width) by 1.2 m (height) to allow for prolonged experimental runtime. The water depth was set at 1.1 m. The double slope configuration was modeled using two heating pads that were insulated at the bottom to avoid heat leakage. The width of the pads was 0.5 m and length along the slope ($L_0$) was 0.3 m.

The desired initial conditions were determined with an aim to isolate the anabatic flow phenomenon from external/mean variations, hence recurring fully stagnant water and uniform temperature field in the tank. Thus, the entire tank was thermally insulated using 5 cm thick polystyrene foam boards. A rectangular section was cut out to enable camera access, and the tank and the cameras were placed inside another insulated enclosure to block out external light and further limit the heat loss. The typical stabilization times, required to achieve the initial conditions, were found to be between 12 and 24 hours, depending on the number of pre-conditions. These included the environmental conditions in the laboratory, whether a simple mixing took place for homogeneous particles seeding or new particles were seeded, or whether a different slope configuration was introduced.

The flow was seeded with neutrally buoyant fluorescent orange polyethylene microspheres with a density of 1.00 g/cc and diameter range of 27-32 µm. This diameter range was selected to ensure that the particles would remain in the field of view during stabilization and prior to heating of the slopes. The particles were illuminated using a continuous 400 mW argon green laser placed in the center of the tank to allow for maximal symmetry. The particle trajectories were recorded using two 1.23-megapixel cameras with a CCD array made up of 1280 horizontal and 960 vertical pixels, with spatial resolution of 40-45 pixels/cm.

Temperature fluctuations were sampled using a grid of 32 k-type thermocouples. Figure 1 depicts the layout of the temperature measurements collected over the right slope. The grid was placed 5 cm behind the laser sheet to avoid obstruction of the flow visualization region of interest.

Data acquisition of flow visualization and temperature records was performed at a sampling frequency of 5 Hz and a recording interval of 4 minutes (1,200 samples) for a total duration of 2 hours (total of 36,000 per experiment). The flow visualization method of choice was Particle Image Velocimetry (PIV). The interrogation window size was 16 pixels by 16 pixels with a 50% overlap. Without subpixel interplation, the minimum trusted velocity is 0.001 m/s. The four-minute interval duration was selected based on the minimal duration required for at least 10 vortices to flow upslope at the slowest speeds examined and limited by the camera-PC-memory link capacity.

We conducted a total of 84 experiments. The conducted experiments ranged in slope angles from $\beta = 5.7°$ to $45°$, thus providing a ratio of apex height to water depth ranging from 0.027 to 0.191. The range of heat fluxes from $Q = 40$ to $1000$ Wm$^{-2}$. Figure 2 provides a comprehensive overview of the distribution of heat fluxes and slope angles. The corresponding buoyancy fluxes were calculated[30] to be in the range of $2 \times 10^{-8}$ m$^2$s$^{-3}$ to $5 \times 10^{-7}$ m$^2$s$^{-3}$. The results presented in the next section are in the slope coordinate system defined in Figure 1, with $s$ denoting the along the slope coordinate, $n$ the normal to the slope axes, and $y$ the transverse (i.e., across the slope). Furthermore, Figure 3 presents the respective slope parallel, perpendicular, and transverse velocity field components, $u, v, w$ respectively, boundary layer height, $\delta$, boundary layer height at the point of separation, $\delta_s$, and estimated along slope length at the point of separation $L_s$.

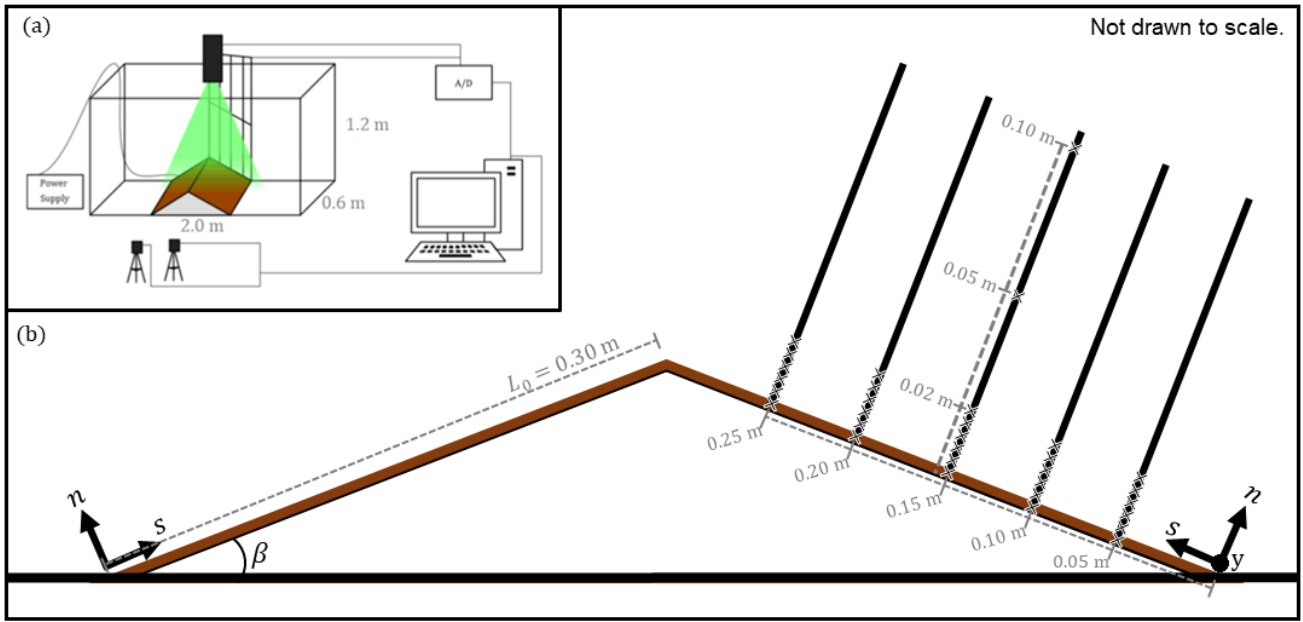

Figure 1 Schematic of experimental apparatus in laboratory. (a) Water tank with a green laser sheet, single and double-sided slopes, cameras, computer, A/D cards, and thermocouples. (b) Grid of K-type thermocouples placed on the center of the slope, with a spacing of 5 cm between each rod and size equally spaced thermocouples spanning 2 cm from the slope, i.e. from 0 cm (on the lope) to 0.02 m, along with two reference points outside the boundary layer, at 5 cm and 10 cm indicated on the middle rod.

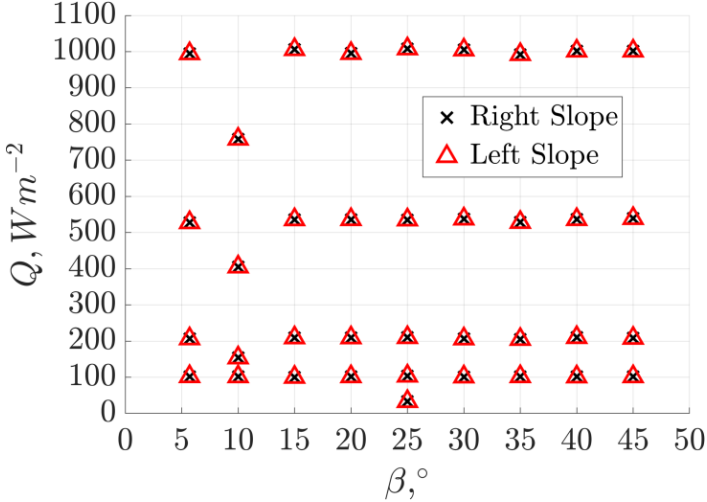

Figure 2 Experimental conditions for all experiments conducted in the water tank with a double slope configuration. The range of angles and heat fluxes are presented, with the differentiation between the slopes indicated by a black ex (x) for the right slope and a red triangle (Δ) for the left slope.

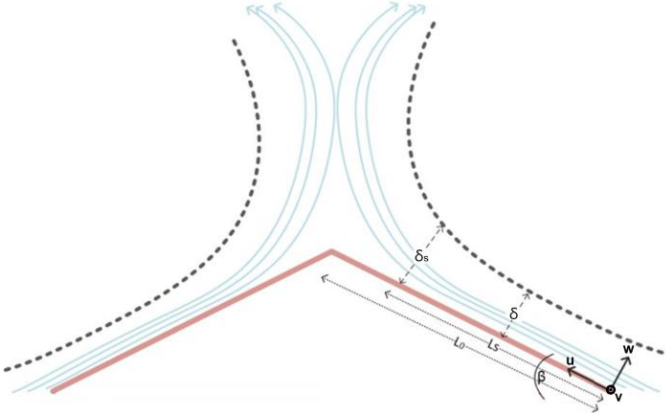

Figure 3 Velocity field components in the along slope coordinate system. The coordinate system is defined separately for each slope, the right slope and left slope. The angle of the slope is $\beta$, slope length is $L_0$, length at the point of separation is $L_s$, height of the BL is $\delta$, height of the BL at separation is $\delta_s$, and the velocity components along slope – u, normal – w, and transverse – v.

## III. RESULTS

Two important nondimensional numbers that characterize anabatic flows are the Richardson number (Ri) and the Reynolds number (Re). The Ri is characterizing the ratio of buoyancy forces to shear forces. In the experiments it was in the range $O(2)$ to $O(10^3)$, indicating dominance of the buoyancy forces relative shear forces, as is expected for a natural convection flow. Since the working fluid in our laboratory experiment was water, we did not expect to achieve Reynolds numbers (Re) comparable to those found in atmospheric flows. The Re measures the ratio of inertial to viscous forces and was found to range from $O(10^1)$ to $O(10^3)$ in our experiments. The Re is important for characterizing anabatic flows on slopes because free convection generates baroclinic torque forces, which drive the flow parallel to the slope, and entrainment is necessary to feed the flow at the leading edge. The Re also incorporates information about the flow geometry, which significantly impacts the flow trajectory. Additionally, the Re is useful for characterizing the shape of the BL. Although Re were not $\gg O(10^3)$, range typically associated with turbulent flow, we did observe that the flow was turbulent in nature. Experimental observations revealed a turbulent flow with vortices rolling along the slope and a plume forming at the apex shortly after the onset of heat flux. Turbulent fluxes dominated the boundary layer, and various intermittencies were observed, as detailed in the following subsections.

### III.a. MEAN BOUNDARY LAYER PROFILES

The convergence of the mean BL profile was assessed during the two-hour experiment by examining four-minute averages of the flow velocity, residual temperature, and turbulent flux profiles.

Velocity profiles were normalized with respect to the maximum along-slope velocity of the 2-hour ensemble, $u_{max}$, resulting in $U_N = u/u_{max}$. We present our results in terms of nondimensional time, $\tau$, defined using the characteristic time scale $\hat{\tau} \equiv \delta/u_{max}$, where $\delta$ denotes the height of the boundary layer at the location of $u_{max}$. Due to the coarse spatial resolution of the velocity field near the boundary, we selected the maximum along-slope velocity instead of the more traditional reference, the friction velocity defined as $u_* = \sqrt{S_{wall}/\rho}$ and representing the effect of fluid stress on the boundary[31]. This is a result of the laser sheet reflectance, which increases the measurement error closer to the boundary. The camera orientation during data acquisition was aligned with the horizontal, and coordinate system rotation was conducted in post processing. For consistency, corrections to the $n$ coordinate were manually made after the rotation of the coordinate system. Future measurements should consider increasing the spatial resolution, as our setup allowed resolving up to four points in the lower logarithmic layer of the velocity profiles. Furthermore, the use of data with low signal-to-noise ratio at only four locations to produce estimates of $u_*$ by a curve fitting method resulted in 95% confidence intervals with errors ranging from $\pm .20 u_*$ to $\pm 800 u_*$. Therefore, it is recommended that future experiments increase the spatial resolution by at least a factor of 2 and consider aligning the laser sheet parallel to the slope.

The temperature profiles were normalized with respect to the temperature on the slope, as $T_N = (T - T_\infty)/(T_w - T_\infty)$, where $T_w$ is the temperature on the slope and $T_\infty$ is the background and initial fluid temperature. Turbulent flux profiles ($\overline{u'w'}$) were calculated using the streamwise and perpendicular velocity field component fluctuations ($u'$ and $w'$, respectively) at each spatial location over the desired duration of either four minutes or two hours. The five points along the slope were selected based on the locations of the available temperature measurements.

We first examine the thirty individual four-minute average profiles for the cases of $\beta$=15°, 25°, and 35° at the highest heat flux of 1000 Wm$^{-2}$. Ensembles of thirty velocity and temperature profiles are presented in Figure 4, and thirty turbulent flux profiles are presented in Figure 5.

The velocity profiles ensemble demonstrates a collapse, indicating the selection of a 4-minute recording interval was sufficient. The collapse within the logarithmic layer has a smaller variance than above, i.e., $(n/\delta) > 1$. The larger variance above the recorded $u_{max}$ suggest increased turbulence levels at $(n/\delta) > 1$. Several profiles exhibit an additional local maximum at $(n/\delta) > 1.5$, making a B-shaped profile, e.g., a profile with more than one inflection points. Examples include $\beta = 15°$ at $L/L_0 = 1/3, 1/2$, $\beta = 25°$ at $L/L_0 = 1/2, 2/3, 5/6$, and $\beta = 15°$ at $L/L_0 = 2/3, 5/6$. The presence of additional bumps above the attached boundary layer suggests the presence of an additional flow structure, which may be due to entrainment or the general shape of the boundary layer (helical structure split discussed later).

The collapse of the non-dimensional temperature profiles in Figure 4 suggests that they exhibit self-similarity and are statistically stationary, especially within the logarithmic layer. The two cases with $\beta = 15°$ and $\beta = 25°$ exhibit nondimensional temperature profiles consistent with inversion conditions. However, the steeper case of $\beta = 35°$ appears to have temperatures at $n/\delta > 1$ that are larger than the temperatures on the slope. The elevated temperatures further from the slope may be ascribed to the sustained heating of vortices within the log layer as they move upstream and may point to reduced entrainment closer to the boundary in the steeper case.

The variance of the nondimensional turbulent flux profiles, $\overline{(u'v')}/u_{max}^2$, tends to increase along the slope. The prominence of the nondimensional turbulent fluxes increases with decreasing $\beta$, suggesting that the flow is more organized at steeper slopes, where the opposing forces of buoyancy and shear converge. Several profiles exhibit an additional local maximum at $(n/\delta) > 0.5$, resulting in a "B-shaped" profile. Examples can be observed at $L/L_0 > 1/6$.

Next, we examine the ensemble-averaged profiles of all experiments over the entire experimental duration. The nondimensional velocity and temperature profiles are shown in Figure 6, and the corresponding nondimensional turbulent fluxes in Figure 7. Figure 6 shows $U_N$ and $T_N$ profiles across all 84 experiments. Additionally, a single slope setup with slope $\beta = 25°$ is presented in the aforementioned figures. The collapse of the $U_N$ profiles across experiments with similar geometries and different sensible heat fluxes confirms that the shape of the boundary layer is independent of thermal forcing but is highly dependent on geometry. The measured velocity magnitudes are, however, positively correlated with heat flux as observed in Refs. [25,26].

Ref. [10] defines three distinct layers within the convective BL on moderate slopes ($\beta < 10°$): the surface layer, the mixed layer, and the inversion layer (see Ref. [10] Fig 2b). The surface layer is the closest to the ground and has a jet-like velocity profile. The mixed layer is above the surface layer and has a uniform velocity profile. The inversion layer is above the mixed layer and has a velocity profile that decreases with height to match the quiescent background conditions. The presented here results from higher slopes also show evidence of the existence of the three layers, while the demarcation between the three layers is more evident in some cases than others, and some of these layers may be absent in certain cases. For example, the mean velocity profile ($U_N$) for a 35° slope at $L/L_0 = 2/3$ exhibits a pronounced surface layer from $0.25 \leq n/\delta \leq 2$. For $q \geq 500 \, \text{Wm}^{-2}$, we observe the absence of a mixed layer but the presence of an inversion layer from $n/\delta > 2$. However, for $q < 500 \text{Wm}^{-2}$, we observe the presence of a mixed layer from $n/\delta > 2$. In the case of the $U_N$ profile for a 25° slope at $L/L_0 = 1/6$, the surface layer is pronounced from $0.5 \leq n/\delta \leq 1.5$, the mixed layer is pronounced $1.5 < n/\delta \leq 3$ and the inversion layer is present from $3 < n/\delta$. The consistent variation in the presence of the modeled layers across all examined cases suggests that the transverse component may not be negligible.

Temperature inversion is observed in all $T_N$ profiles, with an emphasis of a larger inversion gradient at higher heat fluxes. The temperature grid mostly captured the log layer at more moderate slopes $\beta < 30°$ and did not extend into the upper part of the BL. Future experiments should use a finer grid, an infrared camera, or thermochromic particles to obtain more accurate measurements. The collapse of $T_N$ profiles is more pronounced in $15° \leq \beta \leq 25°$ and $\beta_s = 25°$. Outside of this range, the profiles at low and high heat fluxes tend to diverge. These particular cases of lower heat fluxes also demonstrate higher $T_N$ values away from the slope, indicating that vortices flow upslope have had more chance to heat up, and since the flow is generally slower, less entrainment of cooler fluid is mixed into the BL.

Another consistent observation evident in all cases is the presence of a displacement-like layer above the slope, but below the surface layer, at which the velocity is near zero. The displacement layer was not captured by existing models of slope flows, suggesting that it may be due to unresolved near-wall processes.

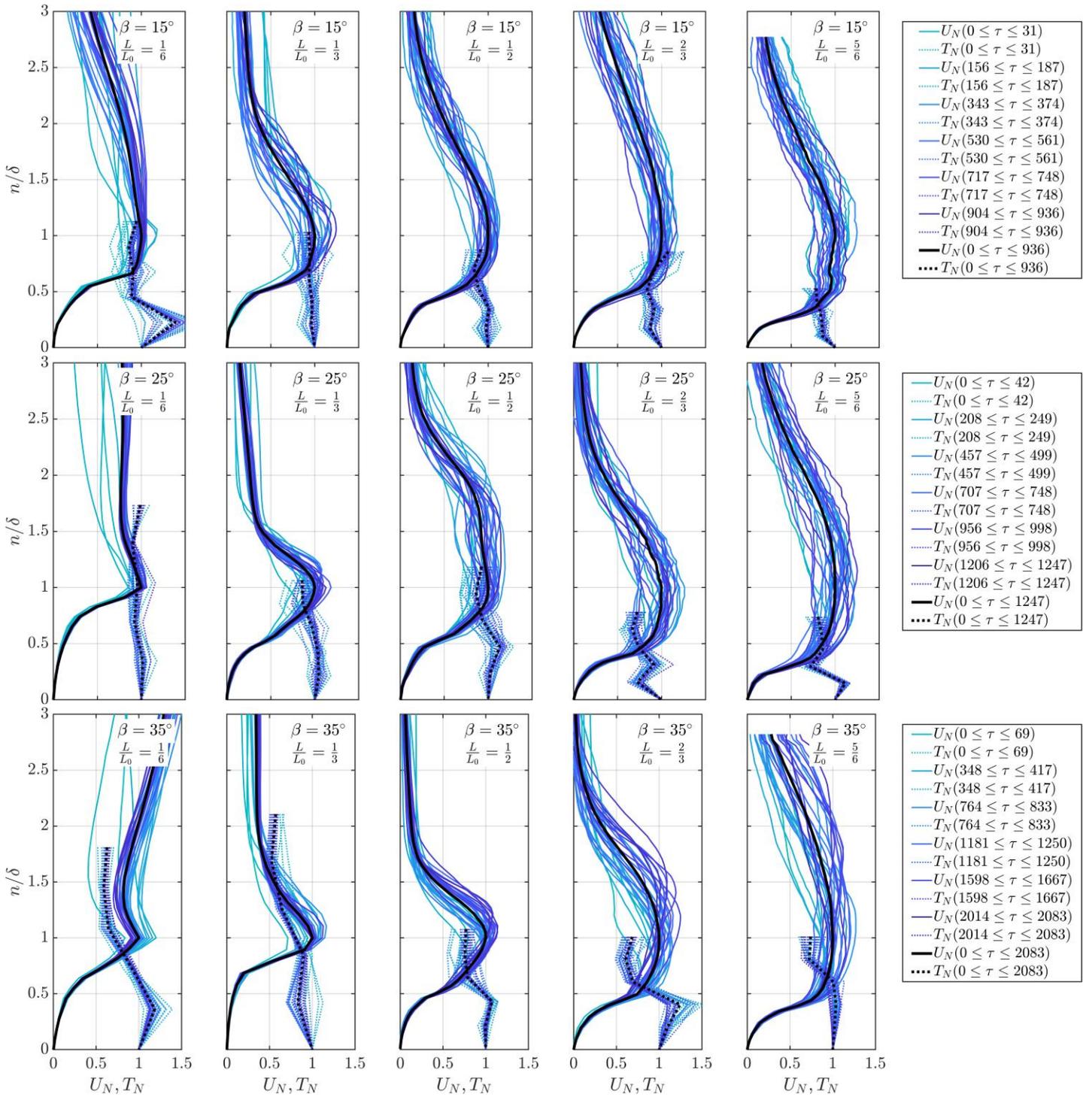

Figure 4 A total of 30 along-slope non-dimensional velocity ($U_N$) and 30 nondimensional temperature ($T_N$) profiles are presented at five cross-sections ($L/L_0$) along the slope. Solid curves represent velocities, dashed curves represent temperatures. These profiles represent the cases with the highest heat flux tested ($Q = 1000$ Wm$^{-2}$). The rows correspond to $\beta = 15°$, $\beta = 25°$, and $\beta = 35°$, from top to bottom. The shade of blue represents the nondimensional time-averaged periods ($\tau$ ranges), ranging from lighter blue shades (beginning of the experiment) to darker blue shades (end of the experiment). Black curves represent the average over the entire experimental duration.

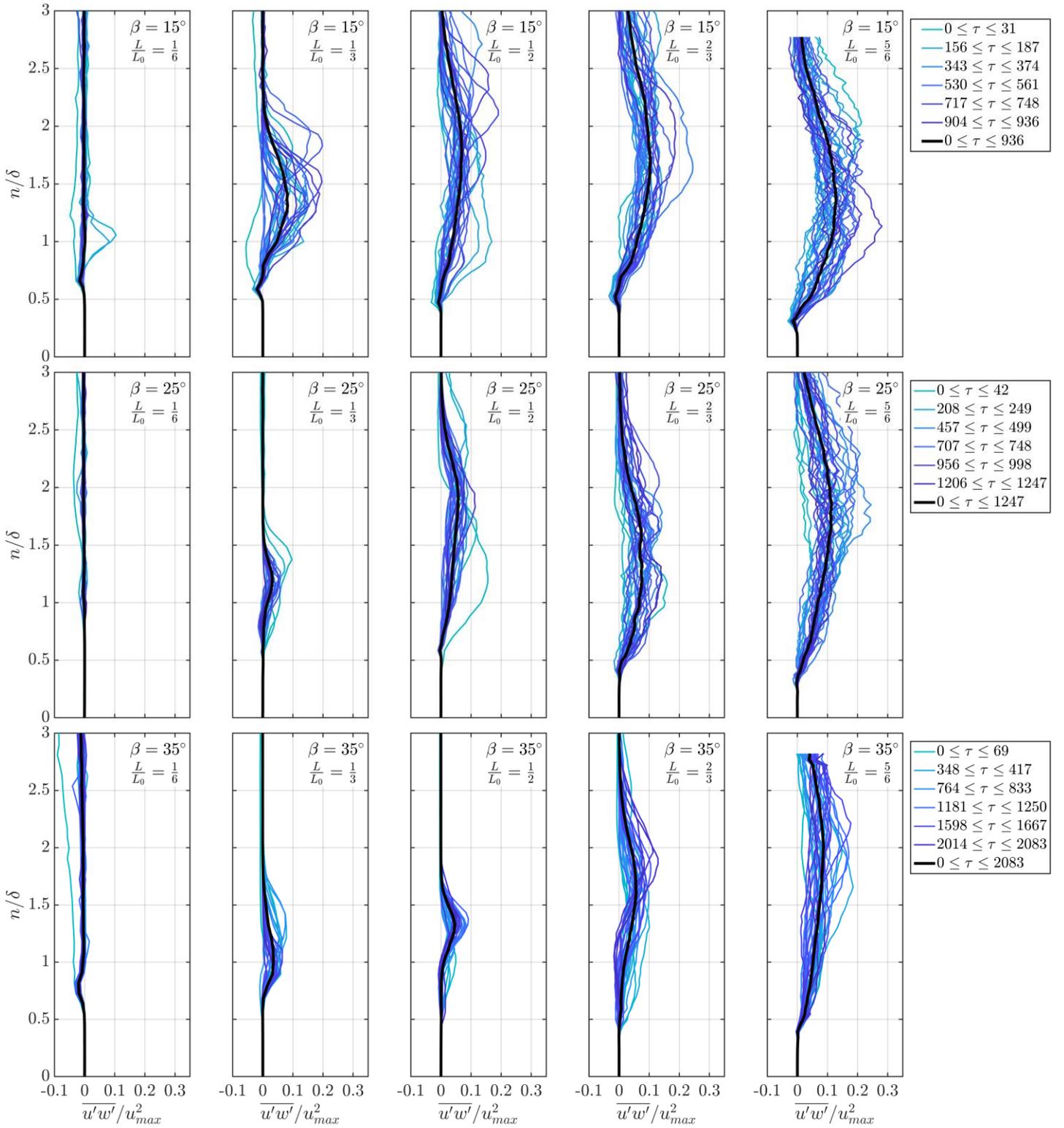

Figure 5 A total of 30 nondimensional turbulent flux ($\overline{u'w'}/u_{max}^2$) profiles are presented at five cross-sections ($L/L_0$) along the slope. These profiles represent the cases with the highest heat flux tested ($Q = 1000$ Wm$^{-2}$). The rows correspond to $\beta = 15°$, $\beta = 25°$, and $\beta = 35°$, from top to bottom. The shade of blue represents the nondimensional time-averaged periods ($\tau$ ranges), ranging from lighter blue shades (beginning of the experiment) to darker blue shades (end of the experiment). Black curves represent the average over the entire experimental duration.

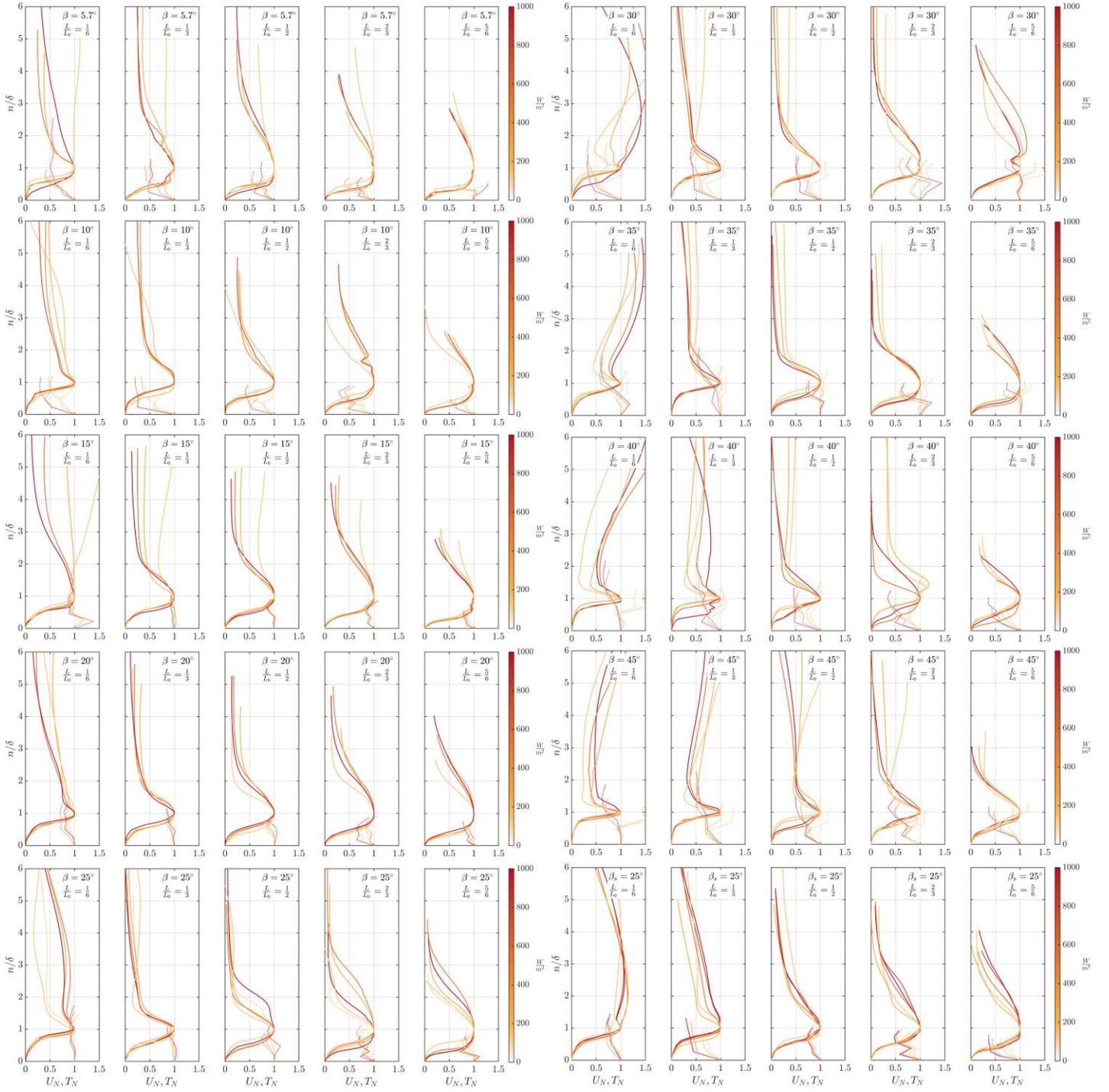

Figure 6 A total of 84 along-slope nondimensional velocity ($U_N$) and 30 nondimensional temperature ($T_N$) profiles are presented at five cross-sections ($L/L_0$) along the slope. Solid curves represent velocities, dashed curves represent temperatures. The colors of the profiles represent the heat flux, which is detailed in the color bar. All double slope configurations correspond to $\beta \in [5.7°, 45°]$ and the single slope configuration corresponds $\beta_s = 25°$. The single slope configuration was created by adding a vertical wall at the apex of the double slope configuration used in the previously discussed experiments. This prevented flow from transferring from one side to the other and were conducted for a direct comparison with Ref. [25].

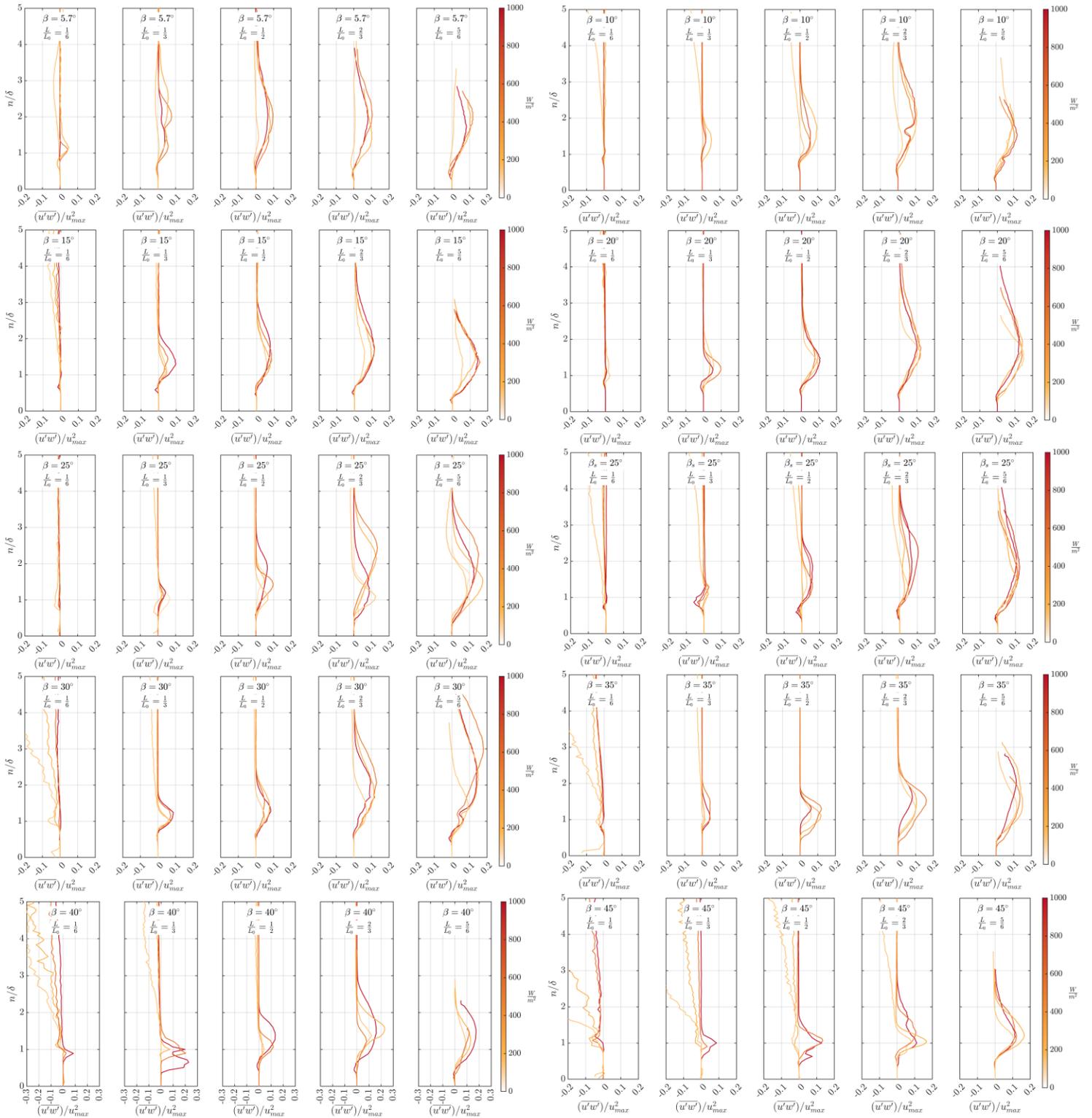

Figure 7 A total of 84 nondimensional turbulent flux ($\overline{u'w'}/u_{max}^2$) profiles are presented at five cross-sections ($L/L_0$) along the slope. Solid curves represent velocities, dashed curves represent temperatures. The colors of the profiles represent the heat flux, which is detailed in the color bar. The double slope configurations correspond to $\beta \in [5.7°, 45°]$ and the single slope configuration corresponds $\beta_s = 25°$. The single slope configuration was created by adding a vertical wall at the apex of the double slope configuration used in the previously discussed experiments. This prevented flow from transferring from one side to the other and were conducted for a direct comparison with Ref. [25].

Future studies with improved resolution near the boundary can explore its origin. We believe that this is not a separation from the slope, as the slope-normal velocity is less than or equal to 0.2 times the mean along-slope velocity. Additionally, the profiles of $\overline{(u'w')}/u_{max}^2$ in Figure 7 exhibit a similar vertical region. This behavior is similar in shape to profiles with zero-plane displacements observed due to obstruction by vegetation or buildings[31–33]. Therefore, the observed behavior suggests that convection may act as an obstruction to the flow. Importantly, near-boundary flow observations confirm parallel, non-separated flow, consistent with the quantitative profiles. However, the possibility of a non-negligible mean transverse flow in this region cannot be ruled out, as it was not measured in this set of experiments.

To further compare our results to the previous studies conducted in a smaller tank [25,26], we conducted an additional set of experiments with the double slope configuration and a vertical wall placed at the apex and between the two slopes that were set up at an angle of $\beta_s = 25°$. This configuration is a physical analogue of the single slope setup that was studied previously[25], albeit in a much larger volume. The normalized velocity and temperature profiles from the "single" slope configuration are presented in Figure 6. The normalized turbulent flux profiles for this single slope configuration of $\beta_s = 25°$ are presented in

Figure 7. The variability between the profiles derived from different heat fluxes is significantly smaller for the single slope configuration than for the double slope configurations. This suggests that the separating wall produces flows that are relatively steadier, as the BLs—and their plumes—on each side of the slope do not interact with each other.

An important observation made in this study is the recurrence of B-shaped mean velocity profiles, characterized by two distinct maxima, in contrast to the expected D-shaped profiles[11,31,34] with one maximum. We hypothesize that the B-shaped velocity profiles correspond to a pair of elongated intertwined helical vortices that move coherently up the slope and, at times, even diagonally up the slope. The observation of helical structures in the flow suggests that the flow is both transient and three-dimensional in nature. This is because helical structures are not typically observed in steady, two-dimensional flows. The B-shaped profile was observed in the four-minute mean profile shapes in Figure 4 in the case of a 15° slope at $L/L_0 = 1/2$, referred to as Evidence 1 (E1). Additional instances were observed in the two-hour $U_N$ average profiles at slopes of 30° at $L/L_0 = 2/3$ as well as in the two-hour $\overline{(u'w')}/u_{max}^2$ average profiles at slopes of 10° at $L/L_0 = 2/3$, which will be referred to as Evidence 2 (E2).

To further examine the instantaneous behavior leading to the B-shaped mean velocity profiles, the behavior of tracer particles in the flow was investigated. Artificial streak videos of the particle flow were generated by taking a sequence of consecutive frames and extracting the brightest pixel at each time step. Producing frames with artificial streak lines of the brightest particles, which is a computationally efficient approach to visualize the flow behavior. Our data repository[35] contains several videos that capture the motion of streak lines, revealing the presence of large lateral movements. These lateral movements provide further evidence (referred to as Evidence 3 (E3)) of the transient nature and three-dimensionality of the flow. Figure 8 shows an example frame in a 25° slope configuration that reveals E3. It demonstrates the splitting of the helical structure and the subsequent separation into two distinct features near the red arrow. The fluid behind the laser sheet at the location of the yellow rectangle would enter the laser sheet and flow along the slope, eventually entering the blue rectangle. The region between the yellow and blue rectangles exhibits a paucity of streak lengths, suggesting either a lack of flow in the observed period or a flow in a plane perpendicular to the field of view. Quantitatively complementing the observations are the profiles of $\partial v/\partial y$ presented in Figure 8(b), which demonstrate the condition $\partial u/\partial s + \partial w/\partial n \neq 0$ is satisfied.

The findings of E1, E2, and E3 demonstrate that the flow is dominantly governed by three-dimensional behavior, which plays a major role in the flow intermittency. This newly observed behavior contradicts a core assumption made in Ref. [25] that the flow is two-dimensional. These findings suggest that the previously developed model should be further re-examined.

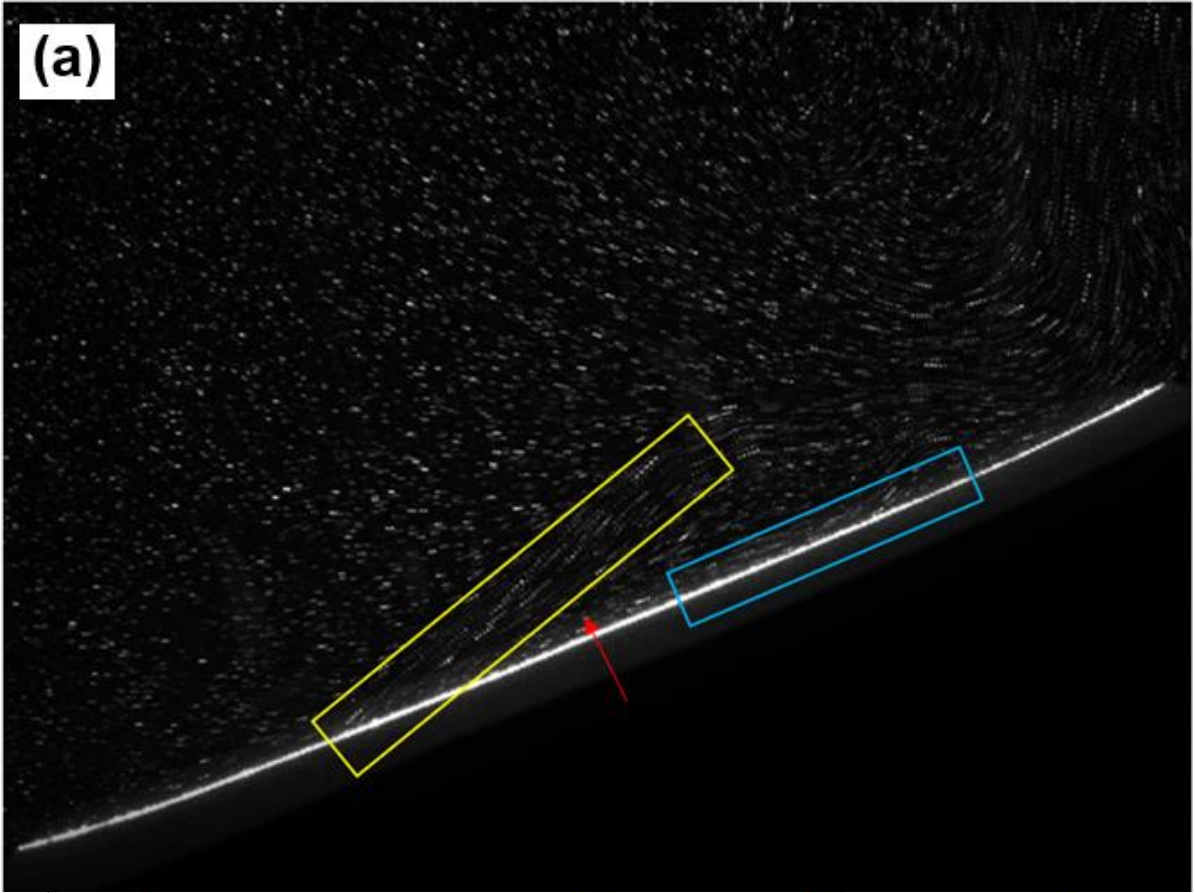

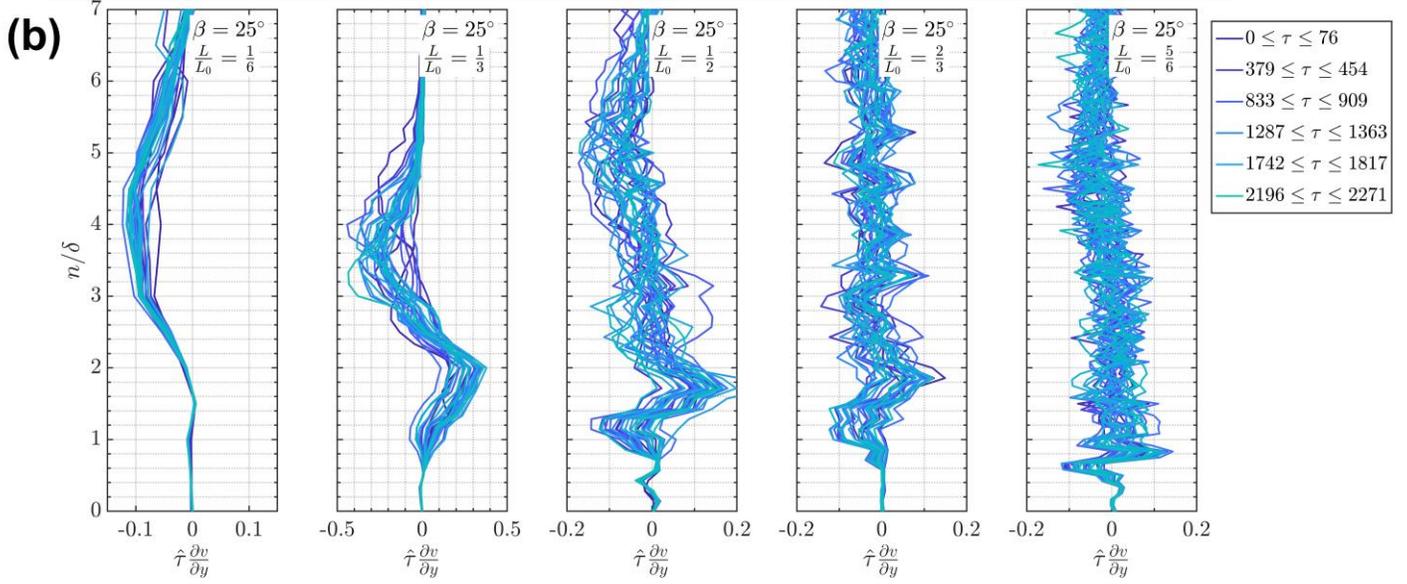

Figure 8 (a) Synthetic streak lines of the flow. The yellow rectangle highlights the helical structure that bifurcates, with the upper part decoupling from the slope and the lower part remaining attached. The blue rectangle highlights the lower part of the structure that remains attached to the slope. The red arrow indicates a region of the flow that is entering the laser sheet from the rear. The complete video can be found in our data repository[35]. (b) Profiles along the slope demonstrating the condition of $\partial u/\partial s + \partial w/\partial n \neq 0$ to provide evidence of the prominence of the transverse component.

## III.b. SEPARATION DETECTION

The next quantifiable step of interest in characterizing the mean boundary layer (BL) behavior is to determine the location at which the flow separates from the slope. Several approaches were considered including an attempt to identify backflow in the mean velocity profiles, as expected from separation of the BL in the literature and presented in Figure 6. However, such behavior was not detected in any of the average velocity profiles examined and may be attributed to either the three-dimensional structure of the flow, the transient nature of the flow, or the buoyancy flux introduced in the experiments. To further investigate this observation, we visually inspected instantaneous velocity vector maps, revealing vortices advecting distally from the slope on most occasions.

Slope-distal advection of vortices was observed in both the mean vorticity and the instantaneous swirling strength maps of the flow, both of which are measures of flow rotationality and used for vortex identification.

Vorticity is a pseudovector field that measures the local spinning motion of a fluid. It is directly related to shear, which is the rate of change of velocity with respect to distance. The vorticity in the y-direction can be estimated[36] as $\omega_y = (\partial w/\partial s) - (\partial u/\partial n)$. Figure 9 presents a sample four-minute average vorticity map of the BL. The surface layer portion of the BL appears to be characterized by two distinct regions. The lower portion (closer to the slope) exhibits consistently negative mean vorticity, suggesting that the flow is attempting to maintain its attachment to the slope. The higher region (immediately above), exhibits positive mean vorticity, suggesting that the flow is decoupling from the boundary-parallel plane or from the vortex structure below it. Examination of the instantaneous vorticity maps reveals the decoupling of the two elongated large structures from each other and will be referred to as Evidence 4 (E4). To confirm that this is indeed the phenomenon occurring along the slope, we next examined vortex identification using swirling strength.

Swirling strength is a measure that estimates the strength of swirling motion, and excludes shear contributions, using the imaginary part of the complex eigenvalue of the velocity gradient tensor[37]. More specifically, it is approximated[38] as

$$\lambda = \frac{1}{2} Im \left[ \frac{\partial u}{\partial s} + \frac{\partial w}{\partial n} - \sqrt{\left(\frac{\partial u}{\partial s} - \frac{\partial w}{\partial n}\right)^2 + 4\left(\frac{\partial u}{\partial n}\frac{\partial w}{\partial s}\right)} \right]. \quad (1)$$

Figure 10 presents a representative instance of identified vortices, with their rotation direction determined by the sign of the vorticity[39]. This map depicts vorticity regions that are primarily influenced by baroclinic torque, rather than shear forcing from entrainment. Positively oriented (red) vortices promote separation and were observed to advect distally from the slope and towards the plume, while negatively oriented (blue) vortices may be associated with entrainment. The experimental data repository[35] consists of a video demonstrating the slope distal advection of vortices.

In view of the observed flow structure, the flow *separation point* was defined as the location along the slope at which the two elongated structures split (as in Figure 9). This is also the point at which vortices, identified using the swirling strength method, were observed to advect distally to the slope. Next, the separation points were manually detected and noted in each four-minute interval. This will be referred to as Evidence 5 (E5). The separation points along the slope were manually identified for all experiments, and the trend is presented in Figure 11. A swarm chart, where points are jittered along the x-axis using a Gaussian kernel density estimate of y weighted by the relative number of points at each x location, is used to represent this trend and demonstrate the distribution for each given $L_s/L_0$ value at each slope angle, $\beta$. A wide range of $L_s/L_0$ values is observed, which necessitates a more informative three-dimensional investigation of the transient flow behavior. Although these distributions show a general increasing trend with increasing angle, the variation at each angle is too large to develop a predictive model using this empirical data. We did not observe a dependence on the heat flux.

To compare our data to the previously developed model[25] for flow separation on smooth slopes, we use the first four minutes of the experiment with the vertical wall ($\beta_s = 25°$) to match the duration of the smaller tank experiments. While Ref. [25] reported an average separation location to be $L_s/L_0 = 0.55$, our observations yielded an average location of $L_s/L_0 = 0.37$, with a range of variations from $L_s/L_0 = 0.17$ to $0.44$. The variation in the separation point locations along the slope is likely due to the three-dimensional transient nature of the flow. This means that the flow is not uniform along the transverse (i.e., across the slope), and it is constantly changing over time. As a result, the location of the separation point may vary depending on the exact location of the laser sheet and the time at which the measurement is taken. The increased size of the tank in our experiments enabled three-dimensional behavior to occur.

We also compared the new geometry to the two geometries previously reported in the literature (a single slope[25] and a single slope with a plateau[26]). We characterized the trends of separation location along the slopes in the double slope configuration without the wall during the first four minutes of each experimental setup, and the results are presented in Figure 12. The first consistent similarity was the general increasing trend of separation location with the slope angle. Two fits were incorporated into the figure to characterize this increasing trend: a linear trend of $0.012\beta$ and the model[26] with the effective angle, $\beta_e = \beta$, with new constants $\Gamma = 395$ and $\Pi = 559$. The nondimensional separation locations are lower than the previous experiments, but they follow a similar increasing trend. Therefore, we conclude that the comparison with the existing model is inadequate because of the dominance of the transient and three-dimensional nature of the flow. The existing model[25] was developed using empirical records from the smaller tank, where the flow behavior was observed to be more organized and two-dimensional. In the larger water tank, the flow has larger variations across the slope, causing an inconsistency with the existing model and the experiments last longer than only the initial development stage of the BL. Another consistent similarity with previous records is the apparent lack of dependence on heat flux. Future development of a more comprehensive model should consider the transient nature and three-dimensionality of the flow. These factors must be incorporated to achieve an accurate representation of the flow and to determine whether quasi-steady conditions of the BL exist in the three-dimensional regime.

### III.c. BOUNDARY LAYER SHAPE

An important conclusion from the previous section is that the flow does not separate from the slope in the conventional sense, but rather that the boundary layer splits into two structures. The upper elongated structure feeds the plume, while the other continues to flow along the slope. The observed shape is similar to classical natural convection flows on the upper surface of an inclined plate[40]. For each of the examined cases, we investigated the average shape of the BL relative to that of a turbulent BL over a flat plate. A reasonable approximation for the thickness of the turbulent BL over a flat (not inclined) plate is given[40] by $\delta_{turb} \approx 0.37xRe_x^{-1/5}$, where $Re_x$ is the local Reynolds number, defined as $Re_x = u_{max}x/\nu$, where $x$ is the distance from the leading edge of the plate and $\nu$ is the kinematic viscosity of the fluid. To fit the empirical height of the turbulent BL in our experiments to the average point of separation, we use the equation $\delta_{BL}(s) = a_0\sqrt{s}$ along the slope coordinate ($s$) until the point of separation or split. Next, we want to compare nondimensional governing parameters to the measured BL shapes.

To compare across different geometries and heat fluxes, we define a shape ratio, $H$, as the ratio of the boundary layer thickness, $\delta_{BL}$, to the turbulent boundary layer thickness, $\delta_{turb}$, for each experiment ($H = \delta_{BL}/\delta_{turb}$). We expect the BL shape to depend on Reynolds (Re) and Prandtl (Pr) numbers. Re measures the ratio of inertial to viscous forces and is defined as $Re = u_{max}s/\nu$, where $s$ is the distance from the leading edge, and $\nu$ is the kinematic viscosity of the fluid. In the context of free convection, Re is important because it contains information about the geometry and shear forcing from entrainment. Since free convection is driven by buoyancy forces, a dependence on the Pr is also expected. Pr is a dimensionless number that characterizes the ratio of momentum diffusivity to thermal diffusivity. These two numbers are therefore useful for predicting the BL shape using several single point measurements within the surface layer. Figure 13a presents the relationship between the shape ratio and the two governing nondimensional numbers. As expected, the shape ratio, $H$, values approach unity as the slope angle, $\beta$, decreases. This indicates that the BL becomes more similar to a turbulent BL over a flat plate as the slope becomes less steep, while it tends to compress as the slope steepens. For most examined cases, except the outlier at $\beta = 5.7°$, the exponent of the Prandtl number in the correlation for the shape ratio of a boundary layer in free convection is approximately -1, and the exponent of the Reynolds number is approximately 0.2. The outlier at $\beta = 5.7°$ was due to the use of only a single point to estimate $\delta_{BL}$ since the average separation was observed at $L_s/L_0 \leq 0.1$.

Finally, the third important parameter for free convection flows is the Richardson number (Ri). The Richardson number is a dimensionless number that characterizes the ratio of buoyancy forces to shear forces. It is defined as $Ri = g\text{B}(T - T_0)\delta/u_{max}^2$, where the thermal expansion coefficient is $\text{B} = 210 \times 10^{-6}\ 1/°C$. Figure 13b shows the relationship between the Richardson number and the Reynolds number for the range of values explored in this study. We observe a logarithmic relationship between the Ri and the Re for steeper slopes ($\beta \geq 25$). More moderate slopes cases, on the other hand, tend to exhibit a constant range of Ri and Re. Additionally, higher heat fluxes tend to correlate with higher Re but lower Ri. Overall, the Ri of atmospheric flows can be matched in laboratory water tank experiments, but the Re cannot. Future laboratory studies should test the same experimental parameters in air to try and match the Re of atmospheric flows.

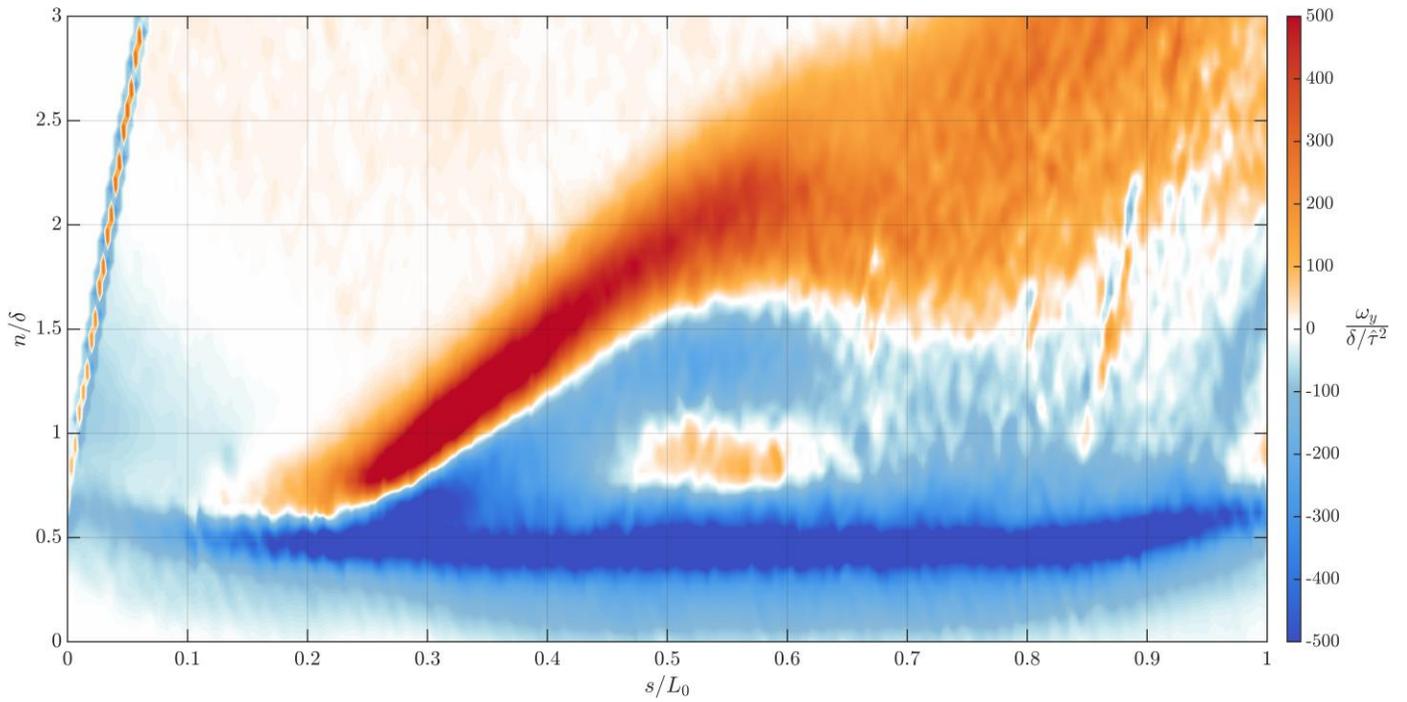

Figure 9 Sample four-minute average ($255 \leq \tau \leq 306$) vorticity map from the double slope configuration at $\beta = 25°$ at a heat flux of $1000$ Wm$^{-2}$. The color scale indicates vorticity, with blue representing negative vorticity, red representing positive vorticity, and white representing near zero vorticity.

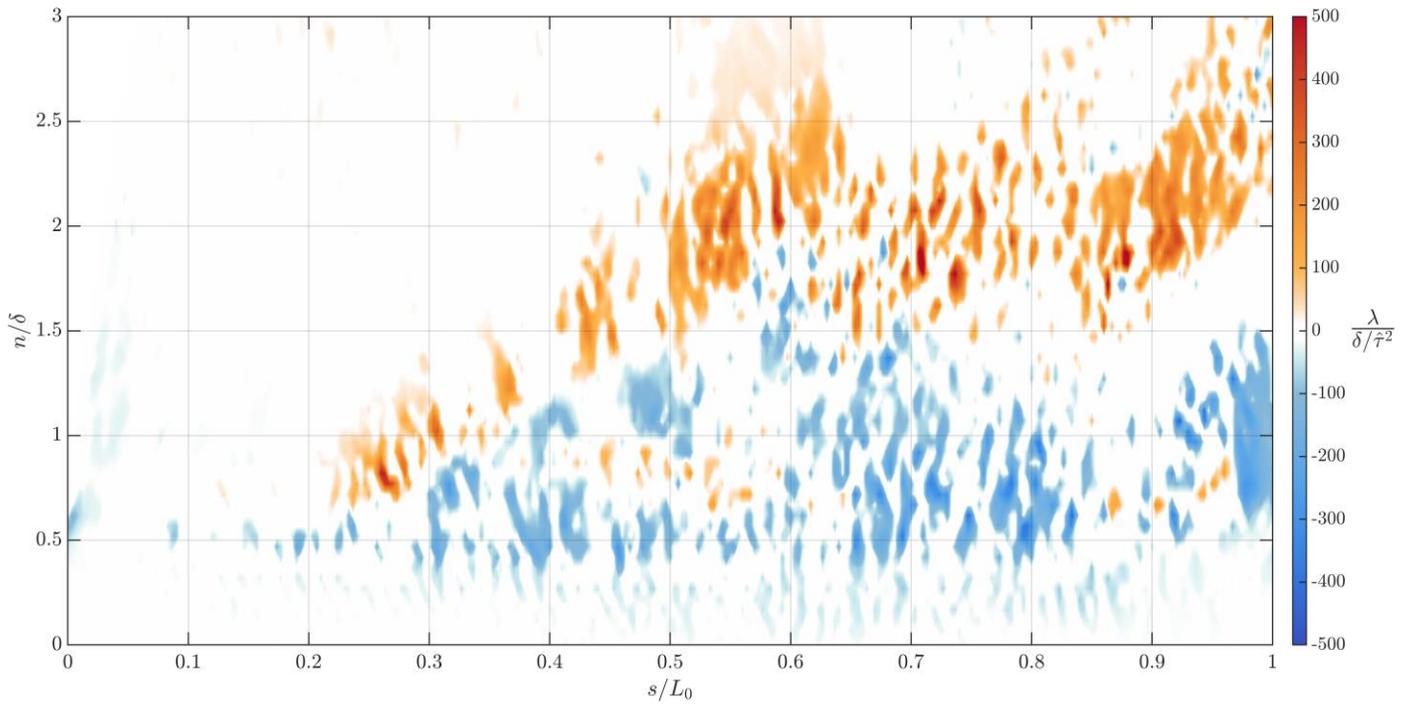

Figure 10 A representative instantaneous ($285 \leq \tau \leq 286$) swirling strength map from the double slope configuration at $\beta = 25°$ at the highest heat flux of $1000$ Wm$^{-2}$. Dark blue regions indicate negative swirling strength, yellow regions indicate positive swirling strength, and light turquoise regions indicate near zero swirling strength (i.e., no detection of a vortex). The signs/directions of rotation for this map were taken from the vorticity maps.

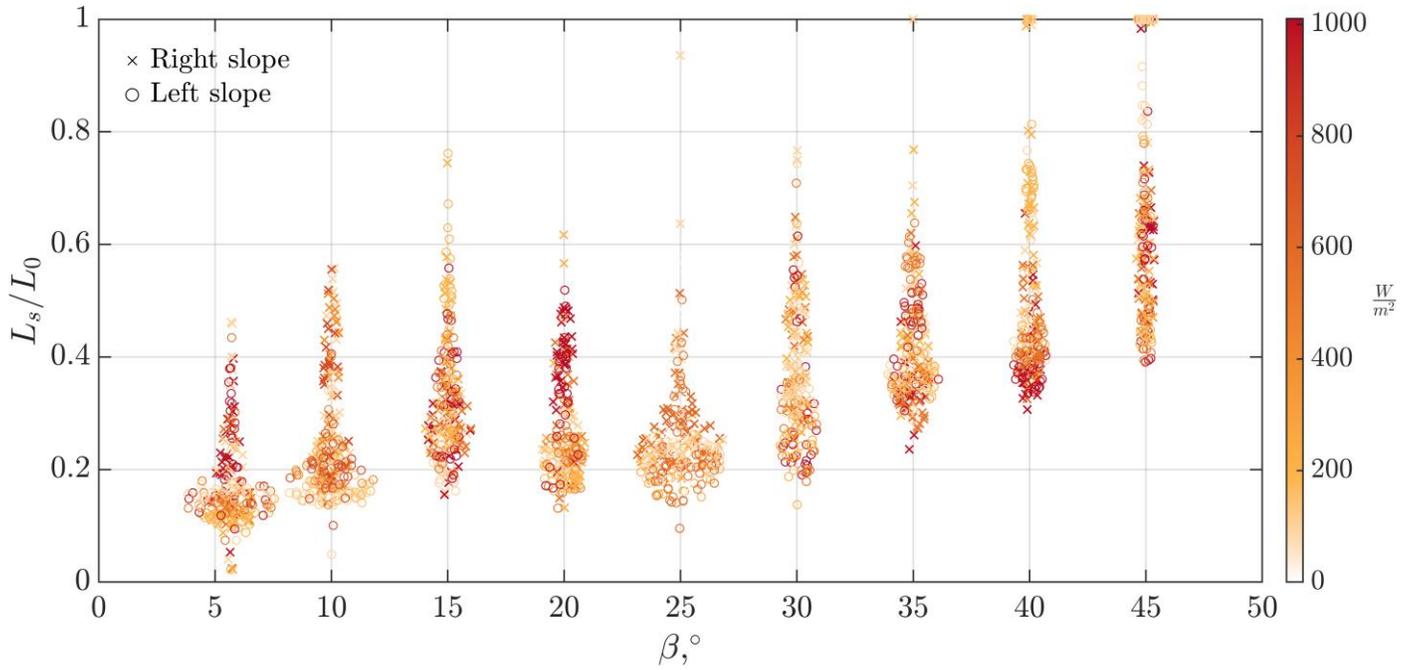

Figure 11 Observed separation location swarm chart distributions for all double slope experiments at all heat fluxes. The colorscale indicates the heatflux. Discrimination between the right and left slopes are noted by x and o markers, respectively.

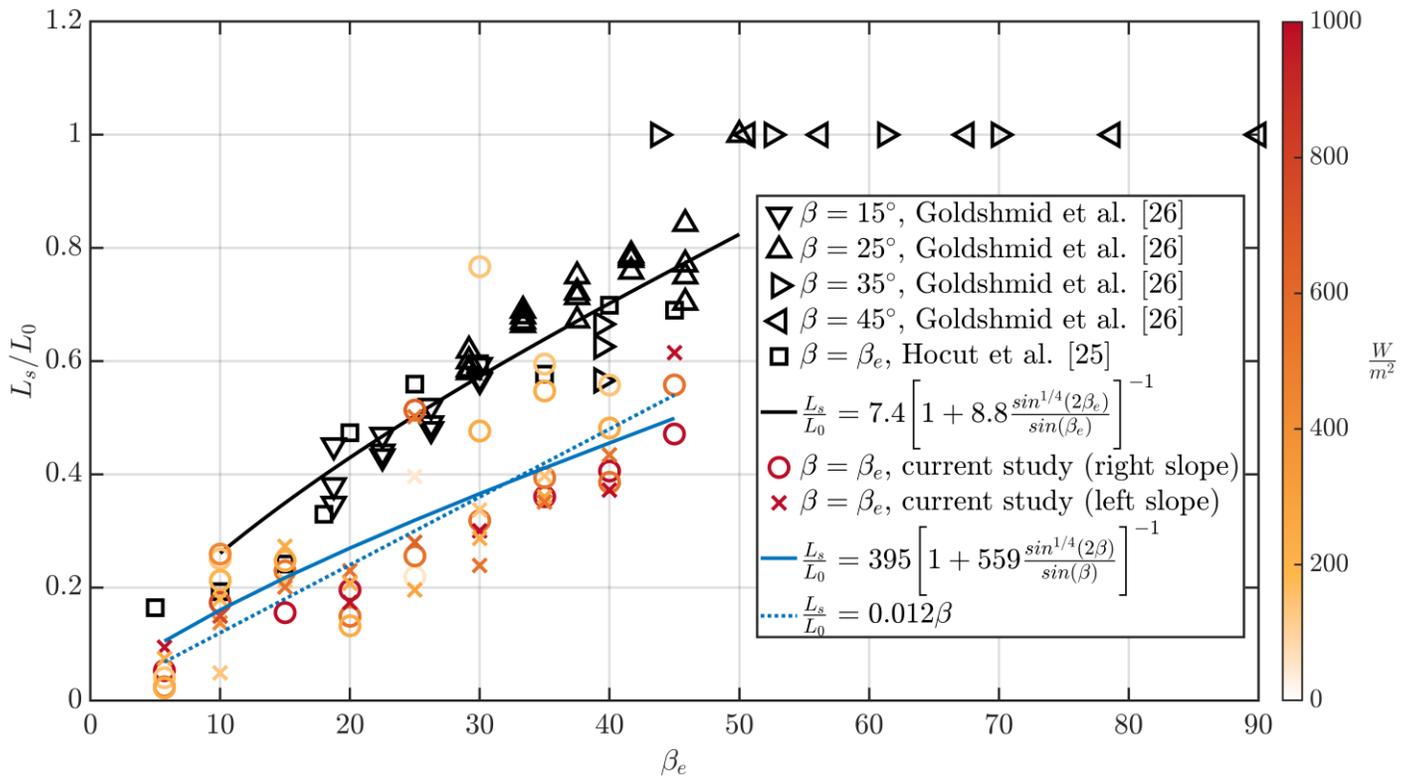

Figure 12 Normalized separation prediction model as a function of effective angle. The observed separation locations for the double slope configuration are overlaid on the adapted figure from Goldshmid RH, Bardoel SL, Hocut CM, Zhong Q, Liberzon D and Fernando HJ, Atmosphere, Vol.9, Article 5, 2018; licensed under a Creative Commons Attribution (CC BY) license. Results from the current study depict the separation locations observed during the first 4 minutes of each experiment of the double slope configuration. The colorscale indicates the heat flux. Discrimination between the right and left slopes is indicated by x and o markers, respectively. The solid blue curve represents the new fit of the previous model with new coefficients, and the dotted blue curve represents a linear fit.

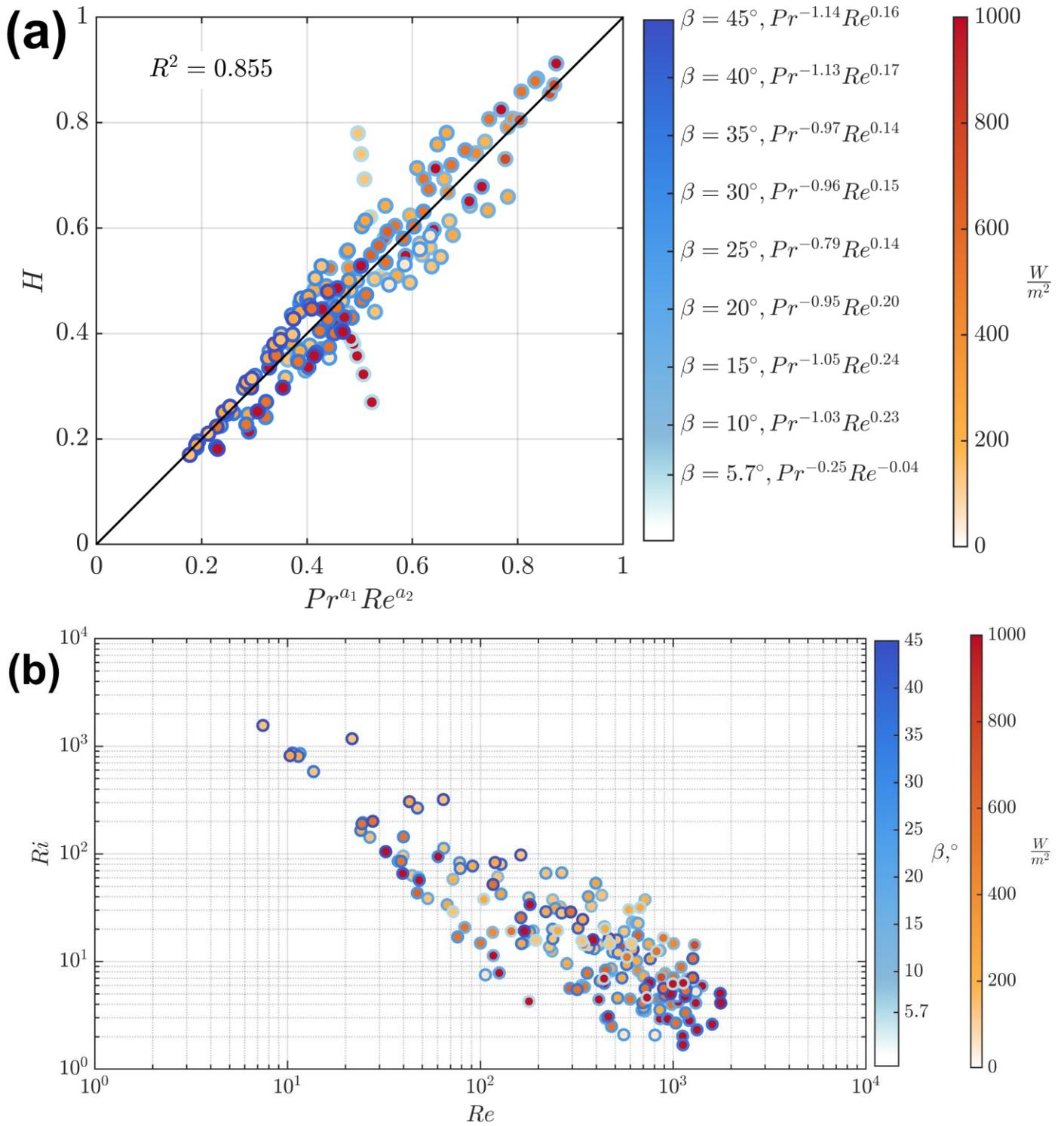

Figure 13 (a) Dependence of the shape ratio on the Prandtl and Reynolds numbers for each experiment. (b) Ranges of Reynolds numbers and Richardson numbers examined in this work.

## III.d. FORCES GOVERNING THE FLOW

Ref.[25] developed a model using the conservation of vorticity equation and the Boussinesq approximation. Under the assumptions of Reynolds similarity, fully developed flow, and 2D quasi-steady flow, the vorticity equation reduced to

$$\frac{\partial \omega_y}{\partial t} = -\frac{\partial(b \cdot cos\beta)}{\partial s} - u\frac{\partial \omega_y}{\partial s}. \quad (2)$$

The y-component of vorticity is denoted by $\omega_y$, and the buoyancy is defined as $b = (\rho_0 - \rho)g/\rho_0$. The baroclinic torque is the process by which density variations induce vorticity, as discussed in Figure 10. Vorticity advection is the process by which shearing forces induce vorticity. In steady state, i.e.,

$$\frac{\partial \omega_y}{\partial t} \equiv 0, \quad (3)$$

the flow was assumed[25] to separate from the slope if the opposing vortices generated by baroclinic torque and vorticity advection were of the same order,

$$-\frac{\partial(b \cdot cos\beta)}{\partial s} \sim u\frac{\partial \omega_y}{\partial s}. \quad (4)$$

The validity of the quasi-steady assumption can be assessed by considering the variation of the dominant forces in the flow.

The thermocouple grid used in our study to obtain the temperature measurements was previously shown to be insufficient to provide a detailed understanding of the temperature and velocity correlations. But the recorded temperature profiles confirm the bulk temperature behaves as expected in an anabatic BL flow. Therefore, an examination of the order of magnitude of the baroclinic torque and vorticity advection along the slope was possible, helping to assess the validity of the quasi-steady assumption, as these forces are the primary drivers of the flow. The spatial separation between the laser sheet and the thermocouple grid precludes direct correlation between the two signals, as this would not yield meaningful information in the along slope dynamics.

We define the nondimensional baroclinic torque as

$$\tilde{B} = -\frac{\partial(b \cdot cos\beta)}{\partial s}\hat{t}^2, \quad (5)$$

and the nondimensional vorticity advection as,

$$\tilde{\Omega} = u\frac{\partial \omega_y}{\partial s}\hat{t}^2, \quad (6)$$

along the slope using the coarse temperature grid. Figure 14 presents average $\tilde{B}$ and $\tilde{\Omega}$ profiles from the entire duration ($0 \leq \tau \leq 1387$) of the experiment for a 25° slope at the highest heat flux ($1000\ Wm^{-2}$) in the regions corresponding to the thermocouple grid layout shown in Figure 1. It is noteworthy that when looking at the time series of $\tilde{B}$ and $\tilde{\Omega}$, they exhibit significant instantaneous fluctuations that vary in time.

The average profiles in Figure 14 demonstrate that the statistically stationary averages of the sum of the two forces diverge from zero. Negative $\tilde{B}$ values are associated with mixing and entrainment-prone conditions and the growth of the BL, while positive $\tilde{B}$ values are associated with separation-prone conditions. Both negative $\tilde{B}$ and near zero values are observed, but we do not observe values of $\tilde{B}$ spanning above the surface layer. If $\tilde{B} \sim \tilde{\Omega}$, the flow is assumed to separate from the slope, and that appears to only occur in the displacement region at $L/L_0 = 1/3$ and $L/L_0 = 1/2$, but these locations were shown to remain attached to the boundary. Therefore, we cannot conclude on the occurrences in the mixed layers and inversion layers until further temperature measurements take place.

Overall, the experimental data presented in this section additionally provides strong evidence that the observed flow is transient and three-dimensional.

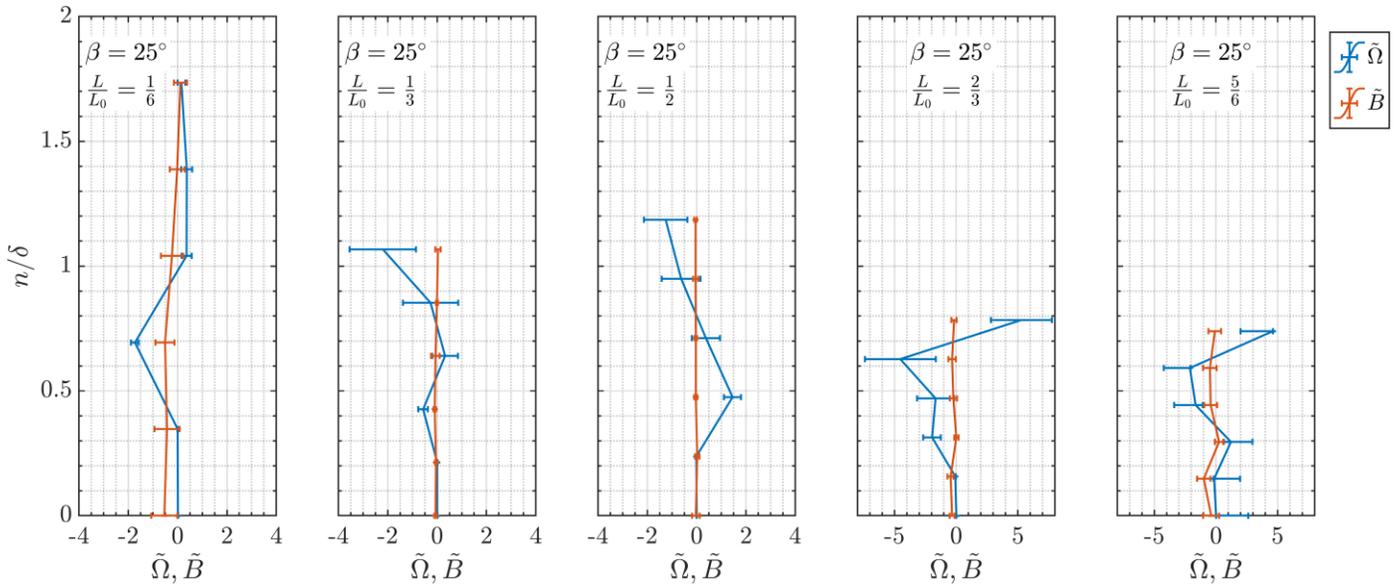

Figure 14  Representative average profiles of the nondimenal baroclinic torque ($\tilde{B}$) and nondimensional vorticity advection in the entire experimental duration $0 \leq \tau \leq 1387$. These profiles correspond to the experiment listed in Figures 9 and 10 above.

### III.e.  THREE-DIMENSIONAL HELICAL STRUCTURES

The previous sections identified five key features (E1-E5) that indicated the presence and dominance of three-dimensionality in the flow. These features motivated a qualitative investigation of the three-dimensionality of the flow in air.

The final set of qualitative experiments was conducted in a thermally insulated polystyrene foam enclosure with dimensions of 1.2 m (length), 0.5 m (width), and 1.45 m (height), using air as the working fluid. The same slope model was used at a slope angle of 10° and a maximum heat flux of 1000 W/m². A Perspex sheet was fitted on one side of the enclosure to provide camera access. A high-volume fog generator was used for flow visualization. Settling chambers were installed at both ends of the enclosure to minimize the initial momentum of the fog.

The flow was observed to consist of helical coherent structures and splitting dynamics. Figure 15 illustrates the hypothesized flow behavior, demonstrating the presence of helical intertwined structures forming and flowing along the slope. In many instances, the coherent structures were also observed to flow diagonally along the slope, see example video in the data repository[35]. This qualitative finding supports the previously proposed hypothesis, and future studies should explore the three-dimensional dynamics of the flow in more detail.

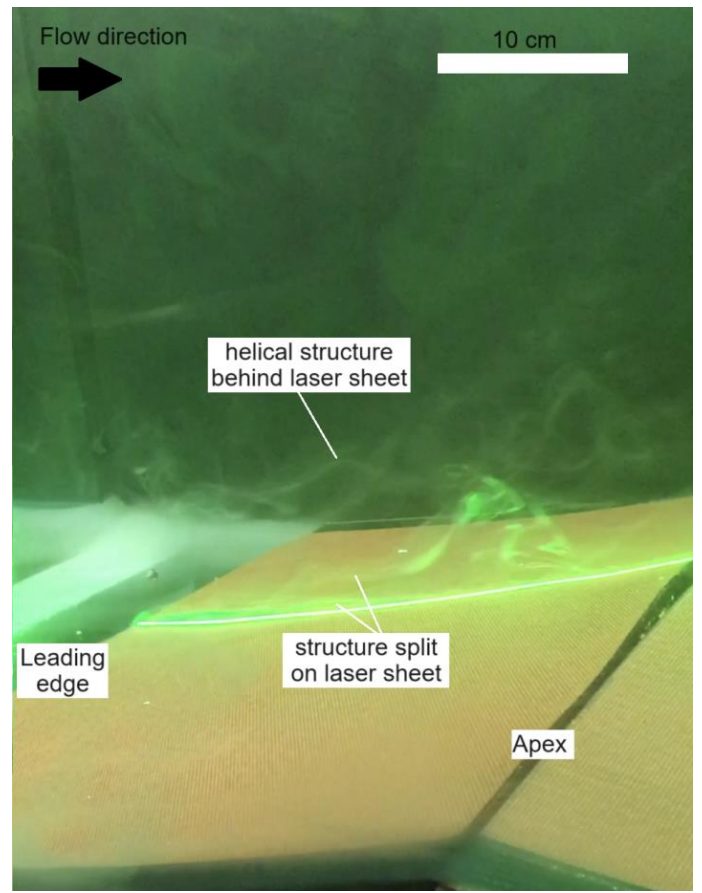

Figure 15  A snapshot of 3D flow visualization over a smooth double slope configuration in air reveals the instantaneous splitting of a structure on the slope, with the lower part remaining attached to the slope and the upper part feeding a plume. A helical vortex structure is also observed behind the laser sheet.

## IV. CONCLUSIONS

This study investigated the behavior of highly turbulent thermally driven anabatic (upslope) flow in a water tank. Particle image velocimetry (PIV) and a thermocouple grid were used to analyze the velocity field and temperature, respectively. The flow exhibited three-dimensional characteristics, with pronounced variations in velocity and temperature. The assumptions of the existing model limited its applicability to purely two-dimensional flow cases, and possibly only to the initial short-time steady period of BL formation along the slope. This was demonstrated by the disagreement of the double slope data with the previous model, which was developed for a single slope[25] and later adapted to one with a plateau [26]. Both of these geometries were previously shown to be approximately two-dimensional within their experimental apparatus, which had a smaller tank and thus limited the experimental duration period. However, the observed disagreement of the double slope data may also be attributed to the increased geometry complexity, which can cause unstable conditions at the apex, and highlight the importance of the three-dimensional nature of the flow.

Five compelling findings (E1-E5) were presented in the results section to underscore the three-dimensional dynamic nature of the flow. The B-shaped mean velocity profiles and turbulent flux profiles (E1 and E2) were only observed when the assumed helical structure intersected the laser sheet. The structures were shown to occasionally flow diagonally along the slope in the 3D visualizations. This clarifies the non-repeatability of observations of the B-shaped profiles. Synthetic streaks (E3) from our data repository[35] reveal particles flowing from behind the laser sheet to the front of it and disappearing. The understanding of 3D behavior using 2D records is incomprehensive and should be properly examined using 3D flow field measurements. Finally, the maps of vorticity and swirling strength (E4 and E5) provided additional evidence of the hypothesized phenomenon of the helical structure split and shooting of vortices from the boundary to the plume. Taken together, these findings demonstrate that the governing flow behavior is three-dimensional and cannot be neglected.

Future work should address the complete three-dimensional nature of the flow. This would provide a better understanding of the anabatic BL behavior and would provide a basis for modeling such flows. The spatial and temporal resolution should also be improved, especially near the boundary, since flow velocity measurements had a small number of points within the logarithmic layer. Improved velocity measurements can be achieved using methods such as 3D flow velocimetry[41–44]. We recommend that future studies would align the laser sheet with the slope and conduct PIV scanning in the direction normal to the slope. Such approach should improve the signal-to-noise ratio near the slope. The temperature measurements can be improved by using a denser thermocouple grid or by using thermochromic particles. This will enable the computation of correlations with the 3D velocity field.

Finally, future studies should introduce surface roughness and additional geometries, such as slope breaks (i.e., changes in slope along the mountain), to better understand the flow dependence on geometry. After that, synoptic-scale modeling should also be introduced to gain a better understanding of the balance between these flows and how synoptic scales influence the boundary layer. Together, these studies can provide novel insights into the flow behavior patterns of complex terrain flows that are driven by thermal forcing.

**Acknowledgement**
This study was funded by the by the Israel Science Foundation (ISF) grant #2063/19.